\documentclass[twocolumn]{aastex63}
\usepackage{amsmath}
\usepackage{graphicx}
\usepackage{grffile}
\usepackage[utf8x]{inputenc}
\usepackage[encapsulated]{CJK}
\usepackage{ucs}
\usepackage[T1]{fontenc}
\newcommand{\cntext}[1]{\begin{CJK}{UTF8}{gbsn}#1\end{CJK}}
\newcommand{\ds}{$\Delta$[\sulf{2}]}
\newcommand{\oiiioii}{[\oxy{3}]/[\oxy{2}]}
\newcommand{\lz}{LzLCS}

\shorttitle{[\sulf{2}]-deficiency and LyC leakage}
\shortauthors{Wang et al.}

\begin{document}
\newcommand{\rom}[1]{{\sc\expandafter{\romannumeral #1\relax}}}
% call as e.g. \si2\; the number is a variable
\newcommand{\sili}[1]{Si~{\rom{#1}}}
\newcommand{\sistar}[1]{Si~{\rom{#1}}$^*$}
\newcommand{\oxy}[1]{O~{\rom{#1}}}
\newcommand{\fe}[1]{Fe~{\rom{#1}}}
\newcommand{\festar}[1]{Fe~{\rom{#1}}$^*$}
\newcommand{\sulf}[1]{S~{\rom{#1}}}
\newcommand{\mg}[1]{Mg~{\rom{#1}}}
\newcommand{\n}[1]{N~{\rom{#1}}}
\newcommand{\nstar}[1]{N~{\rom{#1}}$^*$}
\newcommand{\carb}[1]{C~{\rom{#1}}}
\newcommand{\carbstar}[1]{C~{\rom{#1}}$^*$}
\newcommand{\hy}[1]{H~{\rom{#1}}}

% To be used inside math mode
% Basically replicates of all the commands above, but with the 'm' prefix
\newcommand{\mAA}{\mbox{\normalfont\AA}}
\newcommand{\logsha}{\rm log_{10} [S\,{\textsc{ii}}]/H\alpha}
\newcommand{\logohb}{\rm log_{10} [O\,{\textsc{iii}}]/H\beta}

\newcommand{\mRom}[1]{{\rm \textsc{\expandafter{\romannumeral #1\relax}}}}
% call as e.g. $\msi2$; the number is a variable
\newcommand{\msi}[1]{{\rm Si\,}{\mRom{#1}}}
\newcommand{\msistar}[1]{{\rm Si\,}{\mRom{#1}}^*}
\newcommand{\moxy}[1]{{\rm O\,}{\mRom{#1}}}
\newcommand{\mfe}[1]{{\rm Fe\,}{\mRom{#1}}}
\newcommand{\mfestar}[1]{{\rm Fe\,}{\mRom{#1}}^*}
\newcommand{\msulf}[1]{{\rm S\,}{\mRom{#1}}}

\newcommand{\mha}{{\rm{H}}\alpha}
\newcommand{\mhb}{{\rm{H}}\beta}
\newcommand{\mlya}{{\rm{Ly}}\alpha}
\newcommand{\mlyb}{{\rm{Ly}}\beta}

\newcommand{\mrlya}{R_{\mlya}}
\newcommand{\rlya}{$R_{\mlya}$}
\newcommand{\mewha}{{\rm EW}_{\mha}}
\newcommand{\mewlya}{{\rm EW}_{\mlya}}

\newcommand{\msfrir}{{\rm SFR(IR)}}
\newcommand{\msfrha}{{\rm SFR(\mha)}}
\newcommand{\msfruv}{{\rm SFR(UV)}}
\newcommand{\msfruvir}{{\rm SFR(UV,IR)}}

\newcommand{\mfesc}{f_{\rm esc}}
\newcommand{\mfrel}{f_{\rm esc,rel}}
\newcommand{\mfabs}{f_{\rm esc,abs}}
\newcommand{\mebvmw}{E{\rm(B-V)_{MW}}}
\newcommand{\mebvint}{E{\rm(B-V)_{int}}}

\newcommand{\mmsun}{M_{\odot}}
\newcommand{\mzsolar}{Z_{\odot}}
\newcommand{\mzsubsolar}{Z_{1/7\odot}}
\newcommand{\mmstar}{M_{\star}}

\newcommand{\mfluxerg}{{\rm erg} \, {\rm s}^{-1} \, {\rm cm}^{-2} \, {\rm Hz}^{-1} \, {\rm sr}^{-1}}
\newcommand{\mflux}{{\rm s}^{-1} \, {\rm cm}^{-2} \, {\rm Hz}^{-1} \, {\rm sr}^{-1}}

\newcommand{\ha}{H$\alpha$}
\newcommand{\hb}{H$\beta$}
\newcommand{\ewha}{EW(\ha)}
\newcommand{\ewlya}{EW(\lya)}

\newcommand{\lya}{L\lowercase{y}$\alpha$}
\newcommand{\lyb}{L\lowercase{y}$\beta$}
\newcommand{\nulya}{\nu_{\rm Ly$\alpha$}}

\newcommand{\sfrir}{SFR(IR)}
\newcommand{\sfrha}{SFR(\ha)}
\newcommand{\sfruv}{SFR(UV)}
\newcommand{\sfruvir}{SFR(UV,IR)}

\newcommand{\fesc}{$f_{\rm esc}$}
\newcommand{\frel}{$f_{\rm esc,rel}$}
\newcommand{\fabs}{$f_{\rm esc,abs}$}
\newcommand{\ebvmw}{$E{\rm(B-V)_{MW}}$}
\newcommand{\ebvint}{$E{\rm(B-V)_{int}}$}

\newcommand{\msun}{$M_{\odot}$}
\newcommand{\zsolar}{$Z_{\odot}$}
\newcommand{\zsubsolar}{$Z_{1/7\odot}$}
\newcommand{\mstar}{$M_{\star}$}

\newcommand{\fluxerg}{${\rm erg} \, {\rm s}^{-1} \, {\rm cm}^{-2} \, {\rm Hz}^{-1} \, {\rm sr}^{-1}$}
\newcommand{\flux}{${\rm s}^{-1} \, {\rm cm}^{-2} \, {\rm Hz}^{-1} \, {\rm sr}^{-1}$}

% 21 cm
\newcommand{\rmd}{{\rm d}}
\newcommand{\rme}{{\rm e}}
\newcommand{\rmi}{{\rm i}}

\newcommand{\nh}{n_{\rm H}}
\newcommand{\xhi}{x_{\rm HI}}
\newcommand{\xh}{x_{\rm H}}

\newcommand{\twos}{$2s$}
\newcommand{\wuf}{Wouthuysen-Field}

\newcommand{\ts}{T_{\rm s}}
\newcommand{\tk}{T_{\rm k}}
\newcommand{\tc}{T_{\rm c}}
\newcommand{\tcmb}{T_{\gamma}}
\newcommand{\taugp}{\tau_{\rm GP}}
\newcommand{\tceff}{T_{\rm c}^{\rm eff}}
\newcommand{\sa}{\tilde S_\alpha}
\newcommand{\dtb}{\delta T_{\rm b}}

\newcommand{\xcmb}{x_{\rm CMB}}

\newcommand{\elyac}{\mathcal{E}_{\mlya, \rm c}}
\newcommand{\elyai}{\mathcal{E}_{\mlya, \rm i}}
\newcommand{\elyar}{\mathcal{E}_{\mlya, \rm r}}
\newcommand{\ecmb}{\mathcal{E}_{\rm CMB}}
\newcommand{\ecomp}{\mathcal{E}_{\rm Comp}}

\newcommand{\fcoll}{$f_{\rm coll}$}
\newcommand{\xiion}{$\xi_{\rm ion}$}
\newcommand{\zion}{$\zeta_{\rm ion}$}
\newcommand{\fstar}{$f_\star$}
\newcommand{\fx}{$f_{\rm X}f_{\rm X,h}$}
\newcommand{\tmin}{$T_{\rm min}$}
\newcommand{\massmin}{$M_{\rm min}$}

\newcommand{\mfcoll}{f_{\rm coll}}
\newcommand{\mxiion}{\xi_{\rm ion}}
\newcommand{\mzion}{\zeta_{\rm ion}}
\newcommand{\mfstar}{f_\star}
\newcommand{\mfx}{f_{\rm X}f_{\rm X,h}}
\newcommand{\mtmin}{T_{\rm min}}
\newcommand{\mmassmin}{M_{\rm min}}

\title{The Low-redshift Lyman-continuum Survey: [S \textsc{ii}]-deficiency and the leakage of ionizing radiation}

\correspondingauthor{Bingjie Wang}
\email{bwang@jhu.edu}

\author[0000-0001-9269-5046]{Bingjie Wang (\cntext{王冰洁}\!)}
\affiliation{Department of Physics \& Astronomy, Johns Hopkins University, Baltimore, MD 21218, USA}
\author{Timothy M. Heckman}
\affiliation{Department of Physics \& Astronomy, Johns Hopkins University, Baltimore, MD 21218, USA}
\author{Ricardo Amor\'{i}n}
\affiliation{Instituto de Investigac\'{i}on Multidisciplinar en Ciencia y Tecnolog\'{i}a, Universidad de La Serena, Ra\'{u}l Bitr\'{a}n 1305, La Serena, Chile}
\affiliation{Departamento de Astronom\'{i}a, Universidad de La Serena, La Serena, Chile}
\author{Sanchayeeta Borthakur}
\affiliation{School of Earth \& Space Exploration, Arizona State University, Tempe, AZ 85287, USA}
\author{John Chisholm}
\affiliation{Department of Astronomy, University of Texas at Austin, Austin, TX 78712, USA}
\author{Harry Ferguson}
\affiliation{Space Telescope Science Institute, Baltimore, MD 21218, USA}
\author{Sophia Flury}
\affiliation{Astronomy Department, University of Massachusetts, Amherst, MA 01003, USA}
\author{Mauro Giavalisco}
\affiliation{Astronomy Department, University of Massachusetts, Amherst, MA 01003, USA}
\author{Andrea Grazian}
\affiliation{INAF-Osservatorio Astronomico di Padova, Vicolo dell'Osservatorio 5, I-35122, Padova, Italy}
\author{Matthew Hayes}
\affiliation{Department of Astronomy, University of Stockholm, AlbaNova, Stockholm, Sweden}
\author{Alaina Henry}
\affiliation{Space Telescope Science Institute, Baltimore, MD 21218, USA}
\author{Anne Jaskot}
\affiliation{Astronomy Department, Williams College, Williamstown, MA 01267, USA}
\author{Zhiyuan Ji}
\affiliation{Astronomy Department, University of Massachusetts, Amherst, MA 01003, USA}
\author{Kirill Makan}
\affiliation{Institut f\"{u}r Physik und Astronomie, Universit\"{a}t Potsdam, D-14476 Potsdam, Germany}
\author{Stephan McCandliss}
\affiliation{Department of Physics \& Astronomy, Johns Hopkins University, Baltimore, MD 21218, USA}
\author{M. S. Oey}
\affiliation{Department of Astronomy, University of Michigan, Ann Arbor, MI 48109, USA}
\author{G\"{o}ran \"{O}stlin}
\affiliation{Department of Astronomy, University of Stockholm, AlbaNova, Stockholm, Sweden}
\author{Alberto Saldana-Lopez}
\affiliation{Observatoire de Gen\`{e}ve, Universit\'{e} de Gen\`{e}ve, 1290 Versoix, Switzerland}
\author{Daniel Schaerer}
\affiliation{Observatoire de Gen\`{e}ve, Universit\'{e} de Gen\`{e}ve, 1290 Versoix, Switzerland}
\author{Trinh Thuan}
\affiliation{Astronomy Department, University of Virginia, Charlottesville, VA 22904, USA}
\author{G\'{a}bor Worseck}
\affiliation{Institut f\"{u}r Physik und Astronomie, Universit\"{a}t Potsdam, D-14476 Potsdam, Germany}
\author{Xinfeng Xu}
\affiliation{Department of Physics \& Astronomy, Johns Hopkins University, Baltimore, MD 21218, USA}

\begin{abstract}
The relationship between galaxy characteristics and the reionization of the universe remains elusive, mainly due to the observational difficulty in accessing the Lyman continuum (LyC) at these redshifts. It is thus important to identify low-redshift LyC-leaking galaxies that can be used as laboratories to investigate the physical processes that allow LyC photons to escape. The weakness of the [S~\textsc{ii}] nebular emission lines relative to typical star-forming galaxies has been proposed as a LyC predictor. In this paper, we show that the [S~\textsc{ii}]-deficiency is an effective method to select LyC-leaking candidates using data from the Low-redshift LyC Survey, which has detected flux below the Lyman edge in 35 out of 66 star-forming galaxies with the Cosmic Origins Spectrograph onboard the Hubble Space Telescope. We show that LyC leakers tend to be more [S~\textsc{ii}]-deficient and that the fraction of their detections increases as [S~\textsc{ii}]-deficiency becomes more prominent. Correlational studies suggest that [S~\textsc{ii}]-deficiency complements other LyC diagnostics (such as strong Lyman-$\alpha$ emission and high [O~\textsc{iii}]/[O~\textsc{ii}]). Our results verify an additional technique by which reionization-era galaxies could be studied.
\end{abstract}

\keywords{circumgalactic medium -- extragalactic astronomy -- reionization -- starburst galaxies -- interstellar medium}

\section{Introduction\label{sec:intro}}

The epoch of reionization (EoR), the phase during which the universe transitions from fully neutral to ionized, remains largely unexplored observationally. At the center is the question regarding the sources responsible for the EoR. Deep imaging with the Hubble Space Telescope (HST) indicates that the ultraviolet (UV) luminosity density of early star-forming galaxies is high enough for them to be the best candidates to provide the ionizing photons necessary for reionizing the universe (e.g., \citealt{Bouwens2016}). Unfortunately, since the universe during the EoR is opaque to ionizing photons, direct observations that access the Lyman continuum (LyC) at these redshifts are impossible.

Identifying LyC emitters (LCEs) at low redshifts thus becomes an important step in the investigation of how galaxies could reionize the universe. Over the past decade, the community has invested in a huge effort to identify small samples of LCEs.
Several proxies for LyC escape have been suggested. Strong \lya\ emission is perhaps the most known one, and has been shown to correlate with LyC emission both in individual galaxies at low $z$ \citep{Verhamme2017} and in stacked samples at $z \sim 3$ \citep{Marchi2018,Steidel2018}. However, the absorption due to the neutral intergalactic medium limits its utility at $z \gtrsim 6$. High [\oxy{3}]/[\oxy{2}] flux ratios, which indicate a high ionization state, have been used to select LyC-emitting Green Pea galaxies \citep{Izotov2016a,Jaskot2019}. This class of galaxies constitutes the majority of LCEs in the literature so far. LyC predictors based on UV absorption lines and \mg{2} emission have also been proposed recently \citep{Chisholm2018,Chisholm2020}.

Among those efforts, \cite{Wang2019} (hereafter W19) tested a new diagnostic for LyC leakage, the relative weakness of [\sulf{2}] nebular emission lines (\ds), in a pilot HST program. Significant emerging flux below the Lyman edge was detected in two out of three [\sulf{2}]-deficiency-selected star-forming galaxies at $z\sim0.3$.

In this paper, we explore the connection between LyC leakage and [\sulf{2}]-deficiency with the expanded data set from the Low-redshift LyC Survey (\lz)---a large HST program aiming for a first statistical sample at $z\sim0.3$. The full sample consists of 66 star-forming galaxies, and is described in \cite{lyc_sample}.

The structure of this paper is as follows. In Section~\ref{sec:def}, we begin by reviewing the physical basis for the [\sulf{2}]-deficiency diagnostic and its definition.
In Section~\ref{sec:data}, we summarize the galaxy samples and the relevant analyses.
In Section~\ref{sec:result}, we assess the robustness of the [\sulf{2}]-deficiency test, and compare it to other proposed LyC diagnostics.
In Section~\ref{sec:discuss}, we discuss the implications for the physical properties of LCEs at low- and high-$z$. Finally, we summarize our conclusions in Section~\ref{sec:conclu}.

\section{[\sulf{2}]-deficiency\label{sec:def}}

\begin{figure*}[ht]
\gridline{
    \fig{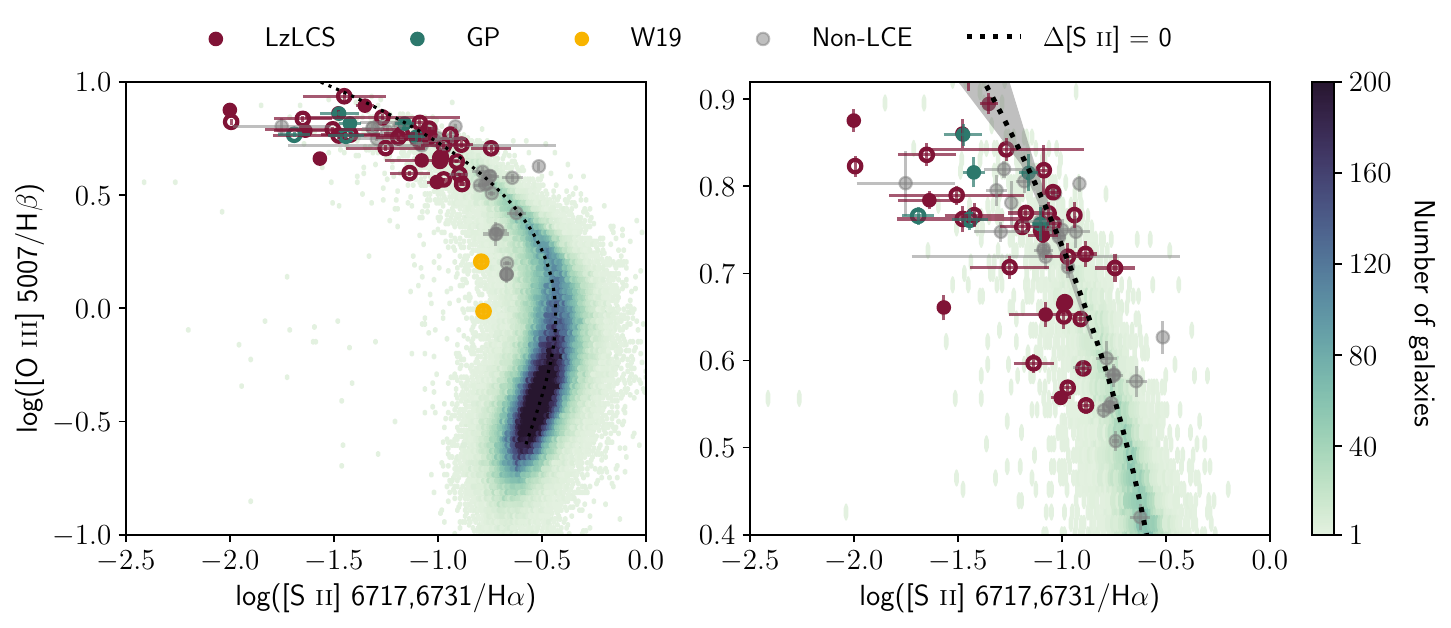}{1.0\textwidth}{}
}
\caption{(Left) This BPT/VO diagram drawn from the SDSS DR12 + BOSS DR8 star-forming galaxy sample is used for defining \ds. The colorbar shows the number of galaxies in each hexbin. The black dotted line is fitted to the locus of the peak density of this distribution. Its uncertainty (gray shade) is negligible expect for log([\oxy{3}]/\hb) $>0.8$, where data becomes sparse. The [\sulf{2}]-deficiency is defined as a galaxy's displacement along the $x$-axis from this ridge line. Also plotted are the galaxy samples considered in this paper: \lz, Green Pea galaxies (GP; \citealt{Izotov2016a,Izotov2016b,Izotov2018a,Izotov2018b}), and [\sulf{2}]-deficiency-selected galaxies (W19; \citealt{Wang2019}). (Right) A zoom-in on the upper part of the left figure. Solid colored dots represent strong LCEs, circles represent weak LCEs, and gray dots are nondetections.\label{fig:bpt}}
\end{figure*}

The relative weakness of the [\sulf{2}] 6717, 6731 emission lines was first proposed as a signpost to identify galaxies that likely allow the escape of LyC radiation by \cite{Alexandroff2015}. This was motivated by a simple physical argument: the ionization potential for producing [\sulf{2}] is only 10.4 eV, which is less than that for ionizing neutral hydrogen. Therefore much of the [\sulf{2}] emission arises in the warm partially ionized region near and just beyond the outer edge of the Str\"{o}mgren sphere in a classical \hy{2} region. This region is weak or even absent when the medium is optically thin to the LyC, resulting in a significant drop in the relative intensity of [\sulf{2}] emission lines \citep{Pellegrini2012}.

We measure the [\sulf{2}]-deficiency in a differential sense: as a quantity relative to the majority of star-forming galaxies in the Sloan Digital Sky Survey (SDSS). 
To do this, we first use the BPT/VO [N~\textsc{ii}]/\ha\ {\emph{vs.}} [\oxy{3}] 5007/\hb\ diagnostic diagram \citep{Baldwin1981,Veilleux1987} to exclude AGN and composite objects, based on the criteria in \cite{Kewley2006}. We then use this sample of star-forming galaxies and the BPT/VO diagram of [\sulf{2}] 6717, 6731/\ha\ {\emph{vs.}} [\oxy{3}] 5007/\hb.
We follow the procedure outlined in W19 in this paper, but with the slight modification of including the star-forming galaxies in the Baryon Oscillation Spectroscopic Survey (BOSS) to improve the sampling of high-excitation galaxies. All fluxes are taken from the value added catalog provided by the Portsmouth group \citep{Portsmouth2013}, and a signal-to-noise (S/N) cut of five is imposed. Both emission-line ratios are dust-extinction corrected based on the observed Balmer lines. Measured values of log([\sulf{2}]/\ha) are binned in log([\oxy{3}]/\hb), and the Gaussian mean (or skewed Gaussian mean when more suitable) of each bin is then calculated.
We define the [\sulf{2}]-deficiency as a galaxy's displacement in log([\sulf{2}]/\ha) from a polynomial fit to Gaussian means. The fitting formula is
\begin{equation}
	y = -0.475 -0.051\xi -0.589\xi^2 -0.360\xi^3,
	\label{eq:ds2}
\end{equation}
where $\xi$ is log([\oxy{3}]/\hb), and $y$ is log([\sulf{2}]/\ha).

We note that the uncertainty in Equation \ref{eq:ds2}, which is estimated via bootstrap, becomes significant only at very large values of log([\oxy{3}]/\hb) > 0.8. Although the difference between the new curve and the one used in W19 or that in \cite{Ramambason2020} is well within $1\sigma$ even for log([\oxy{3}]/\hb) > 0.8, we have tested the result of excluding galaxies lying above this value on the subsequent analysis in the paper, and find that there are no changes to our conclusions.

One other potential issue is that while the ratio of [\oxy{3}]/\hb\ generally increases with decreasing O/H, it reaches its maximum value at 12 + 1og O/H $\sim 7.9$, below which it slowly falls \citep{Maiolino2008}. For galaxies below this metallicity value, a decrease in [\oxy{3}]/\hb\ could potentially mimic a [\sulf{2}] deficiency in Figure \ref{fig:bpt}. To check on the potential effects of this we have also tested the result of excluding the 10 sample members with $12 + {\rm log O/H} \leq 7.9$. We again find that there are no changes to our conclusions.

The samples of galaxies are shown as filled and unfilled circles in the Figure~\ref{fig:bpt} and their relevant analyses are presented in Section \ref{sec:data}.

\section{Data\label{sec:data}}
\subsection{Galaxy samples}

\lz\ (HST-GO-15626; PI A. Jaskot) consists of a sample of 66 star-forming galaxies at $z \sim 0.3$, selected to meet at least one of the following three criteria: [\oxy{3}]/[\oxy{2}] flux ratio $>3$, UV spectral slope $\beta < -2$, or $\Sigma_{\rm SFR} > 0.1 \, \mmsun {\rm yr}^{-1} {\rm kpc}^{-2}$.

We additionally consider the following two samples from the literature.
First, three star-forming galaxies were selected based on [\sulf{2}]-deficiency in a pilot program (HST-GO-15341; PI T. Heckman), two of which have been observed with significant LyC flux in W19.
Second, 11 Green Pea galaxies (GPs) are included, which constitute the majority of confirmed low-$z$ LCEs in the existing literature before \lz\ \citep{Izotov2016a,Izotov2016b,Izotov2018a,Izotov2018b}.

\subsection{Data analysis\label{sec:data_analysis}}

Processing of the Cosmic Origins Spectrograph (COS) spectra in \lz\ is presented in detail in \cite{lyc_sample}. UV continuum fitting is discussed in \cite{lyc_uvfit}. Here we describe the additional analyses performed for the purpose of this paper.

The fluxes of [\sulf{2}] emission lines in SDSS spectra of are remeasured for all sources in \lz. This is done out of caution as some \lz\ spectra are found to have low S/N in the [\sulf{2}] lines. However, this does not imply that the fluxes of the SDSS reference sample used to draw the curve of \ds\ = 0 are subject to a similar error, since only emission lines with S/N $\geq$ 5 are selected.

For SDSS spectra of the \lz\ sample, we categorize them into three subgroups based on S/N and each is subjected to a different treatment: 1) both lines of the [\sulf{2}] doublet, when detected, are simultaneously fitted with Gaussians, and $1\sigma$ uncertainties as provided in SDSS spectra are propagated; 2) when only one of the [\sulf{2}] lines is detected with significance, the undetected line is inferred from the detection; that is, its Gaussian fit is constrained by the line-center shift, amplitude (after taking the typical ratio of the doublets into account), and the full width at half maximum of the detected line; 3) when neither of the [\sulf{2}] lines are available, $3\sigma$ upper limits are inferred from uncertainties. We list the measured flux along with [\sulf{2}]-deficiency in the Appendix.

\subsection{Escape fractions\label{sec:data_fesc}}

Given the complications in estimating escape fractions (\fesc) of the LyC photons, two approaches are considered. First, the ratio between the flux of LyC and of the stellar continuum at $\sim 1100$~\AA\ rest frame ($f_{\rm LyC}/f_{1100}$) is used as a proxy of \fesc. It has the advantage of being less model dependent. 
Second, it is common to estimate \fesc\ by comparing the observed ratio of flux density at $\sim 900$~\AA\ to that at $\sim 1500$~\AA\ ($F_{900}/F_{1500}$) with the intrinsic ratio \citep{Steidel2001}. After accounting for the dust attenuation, we obtain the absolute \fesc, denoted as $f_{\rm esc}^{\rm UV}$(LyC) in this paper. Specifically, we derive $f_{\rm esc}^{\rm UV}$(LyC) by finding an intrinsic UV spectrum following the fitting methods of \cite{Chisholm2019} and presented in \cite{lyc_uvfit}. We fit the observed stellar continuum as a linear combination of multiple single-age and single-stellar-metallicity bursts. We use 50 possible theoretical Starburst99 models \citep{Leitherer1999, Leitherer2010} that span a range of ages (1 -- 40 Myr) and metallicities (0.05 -- 2 $Z_\odot$) that are relevant to the young starbursts in the \lz. We assume a standard Kroupa initial mass function \citep{Kroupa2002} and use the Geneva stellar evolution tracks with high-mass-loss rates \citep{Meynet1994}. Finally, we account for dust attenuation using the \citet{Reddy2016} law, which is observationally defined down to 950~\AA. We fit to the stellar continuum redward of 950~\AA\ by masking the strong interstellar medium (ISM) absorption features and fitting the best-fit linear coefficients and dust attenuation parameter. We then extend the stellar continuum fit blueward to LyC and take the ratio of the fit to the observations to determine $f_{\rm esc}^{\rm UV}$(LyC).

\section{Results\label{sec:result}}

\subsection{Statistical tests of the {[S II]}-deficiency diagnostic\label{sec:ds2_leak}}

\begin{figure}
\gridline{
    \fig{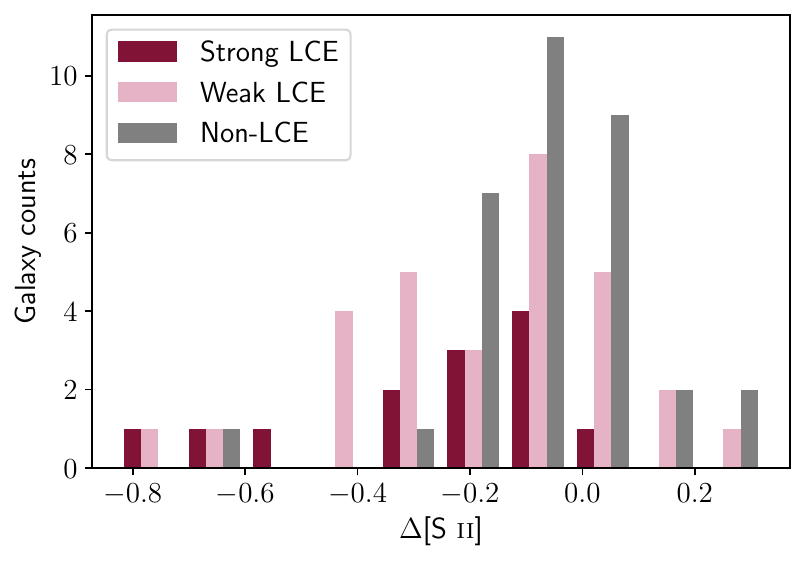}{0.47\textwidth}{(a)}
}
\gridline{
    \fig{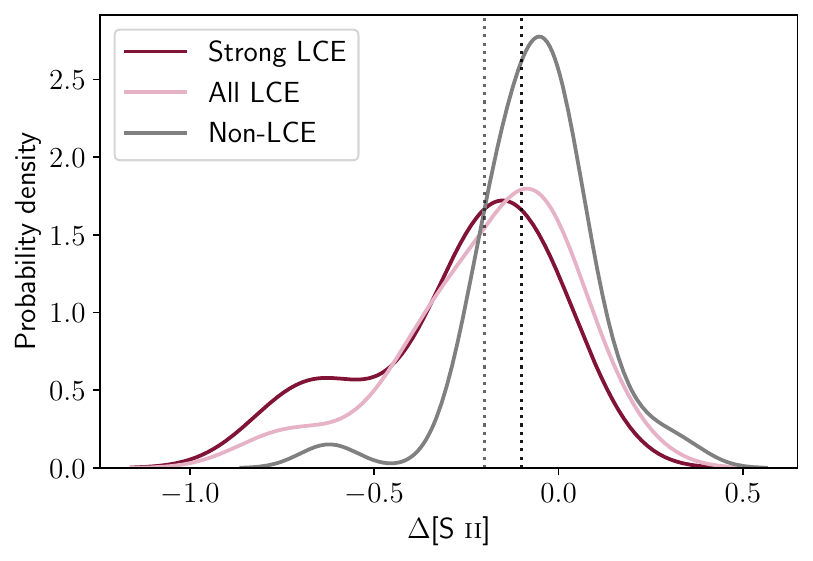}{0.47\textwidth}{(b)}
}
\gridline{
    \fig{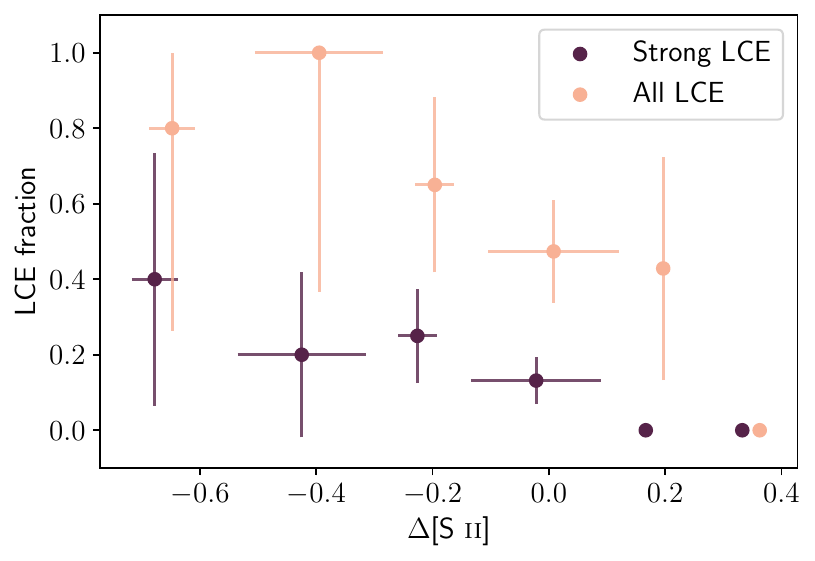}{0.47\textwidth}{(c)}
}
\caption{(a) Histograms and (b) Gaussian KDEs showing distributions of \ds\ among the whole sample. The two dotted lines represent \ds\ = -0.2 and -0.1 respectively. (c) Fractions of strong and all LCE detections in bins of \ds. Results from different samples are offset in \ds\ for clarity. Correlations are exhibited despite of substantial Poisson uncertainties driven by small number counts  (see Table \ref{tab:r}). \label{fig:res_ds2}}
\end{figure}

\begin{deluxetable}{lcc}
\tablecolumns{3}
\tablewidth{0pc}
\tablecaption{Correlations between \ds\ and \fesc / LCE Fractions\label{tab:r}}
\tablehead{
    \colhead{} &
    \colhead{Kendall's $\tau$} &
    \colhead{$p$-value}
    }
\startdata
$f_{\rm LyC}/f_{1100}$\tablenotemark{a} & $-0.256$ & $0.001$ \\
$f_{\rm esc}^{\rm UV}$(LyC)\tablenotemark{a} & $-0.229$ & $0.003$ \\
\hline
$\mathit{F}$(s)\tablenotemark{b} & $-0.828$ & $0.022$ \\
$\mathit{F}$(all)\tablenotemark{c} & $-0.867$ & $0.017$ \\
\enddata
\tablenotetext{a}{Correlations are calculated from the whole sample (see Figure \ref{fig:fesc}).}
\tablenotetext{b}{Fractions of strong LCEs in bins of \ds\ (see Figure \ref{fig:res_ds2}c).}
\tablenotetext{c}{Fractions of all LCEs in bins of \ds.}
%\tablecomments{Uncertainties are determined via bootstrap.}
\end{deluxetable}

The larger sample size of \lz\ allows for a first statistical study on the reliability of the [\sulf{2}]-deficiency diagnostic. Before proceeding to present the results, we would first like to reiterate definitions of LCEs in the survey \citep{lyc_sample}.
The classification is determined based on two criteria: the probability ($P(> N|B)$) that the observed or gross counts within the extraction window of the LyC are due to background fluctuations \citep{Worseck2016}, and escape fraction (\fesc) or its proxy $f_{\rm LyC}/f_{1100}$.
 First, we define LCEs as having $P(> N|B) < 0.02275$. Second, a subset of LCEs having $P(> N|B) < 2.867 \times 10^{-7}$ and \fesc\ or $f_{\rm LyC}/f_{1100} > 0.05$ are defined as ``strong'' detections. The rest are considered non-LCEs. We note that using different definitions of \fesc\ yields qualitatively similar results.

We address the robustness of the [\sulf{2}]-deficiency diagnostic from two perspectives.
First, distributions of \ds\ among the different groups are compared. Figure \ref{fig:res_ds2}(a)-(b) show the histograms and Gaussian kernel density estimates (KDEs) of \ds, respectively. A preference for the class of LCEs to have more negative \ds\ than that of the non-LCEs is present.

To quantify the above, we calculate an Anderson-Darling statistic for the LCEs (and strong LCEs) of 3.0 (2.3), suggesting that the null hypothesis that the two samples come from the same distribution can be rejected at about a $98.0\%$ ($96.3\%$) level.

Second, we calculate fractions of LCE detections in bins of \ds, as shown in Figure \ref{fig:res_ds2}(c). Significant correlations are exhibited, despite substantial Poisson uncertainties on fractions which are driven by the small number counts. We list the correlation coefficients in Table \ref{tab:r}. 

Taken together, those results indicate that a candidate's likelihood of being an LCE increases as [\sulf{2}]-deficiency becomes more prominent.

\subsection{Comparison to other LyC diagnostics\label{sec:twin}}

\begin{figure*}[ht]
\gridline{
    \fig{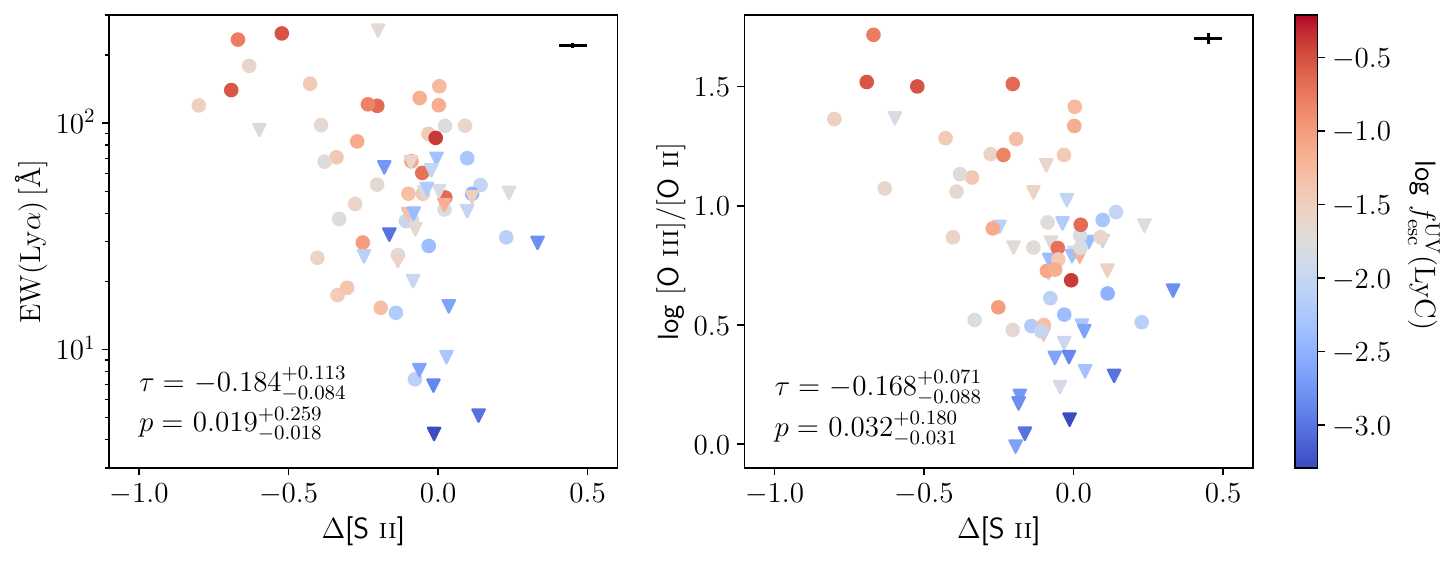}{1.0\textwidth}{}
}
\caption{Scatter plots showing correlations between \ds\ and other galaxy characteristics. Colors indicate \fesc. Triangles indicate that the associated \fesc\ are upper limits. $\tau$ and $p$ are Kendall's correlation coefficient and the $p$-value, respectively. All show substantial scatter; the rest of the plots can be found in the Appendix.\label{fig:ds2_cor}}
\end{figure*}

A number of galaxy characteristics have been identified as potential signposts of LyC leakage (e.g., \citealt{Verhamme2015,Izotov2016a,Chisholm2018}; for results from \lz, see \citealt{lyc_fesc}).
Here we evaluate the [\sulf{2}]-deficiency diagnostic in context of its relationship to other signposts.
We note that the velocity separation ($v_{\rm sep}$) of the two Ly$\alpha$ peaks and \fesc\ of the LyC has been shown to be tightly correlated \citep{Izotov2018b,Izotov2021}. However, \lz\ lacks the necessary observations using the low-resolution COS/G140L grating to measure $v_{\rm sep}$. We hence neglect a discussion regarding $v_{\rm sep}$ in what follows.

Figure \ref{fig:ds2_cor} shows the galaxy distributions in the plane of \ds\ \emph{vs.} other signposts, color coded in \fesc\ (see Section \ref{sec:fesc} for details on \fesc). The corresponding Kendall's $\tau$ correlation coefficients and their significance ($p$-values) are also shown in the figures. Their uncertainties are estimated via bootstrap.
Most of the correlations are rather weak. Among the strongest ones are between \ds\ and the half-light radii ($r_{50}$) measured from COS NUV ACQ images and the star formation rate per unit area ($\Sigma_{\rm SFR,H\beta}$). This is consistent with the frequent identification of LCEs as being highly compact. It is then reasonable to speculate that LyC escape can be made possible by the extreme feedback effects produced by a ``dominant central object.'' These objects are defined to be compact, very massive, and young objects located at or near the galactic nucleus \citep{Heckman2011a,Borthakur2014}. Similar findings have likewise been reported in \cite{Wang2019,Hogarth2020,Kim2020,Kim2021}.

Although adopting the standard discriminant of $p=0.05$ means that most of the correlations are not significant, we do see significant additional correlations with EW($H\beta$), $M_{\rm FUV}$, $M_{\rm NUV}$, [\oxy{3}]/[\oxy{2}], and SFR$_{\rm UV}$/$M_\star$ In all cases, the scatter is substantial. This suggests that [\sulf{2}]-deficiency is providing information on LyC leakage that is largely independent of the other signposts.

To further demonstrate this, we construct pairs of [\sulf{2}]-weak (\ds\ $\leq -0.2$) and non-[\sulf{2}]-weak galaxies (``twins'') that share the closest values for each of the other parameters shown in Figure \ref{fig:ds2_cor} and in the Appendix. For example, in the case of the EW(Ly$\alpha$) parameter, for each [\sulf{2}]-weak galaxy, we find a non-[\sulf{2}]-weak galaxy that is the closest match in EW(Ly$\alpha$). When one galaxy has multiple twins, we remove the duplicates from calculations. We then compare LCE fractions between the two subsamples and test the statistical significance of the difference. The results are listed in Table \ref{tab:lyc_frac}. We note that LCE fractions in the [\sulf{2}]-weak sample are greater than those in the twin samples in all cases, although for some parameters the large uncertainties due to the relatively small sample size lead to statistically insignificant results.

\begin{deluxetable*}{lcccc}
\tablecolumns{5}
\tablewidth{0pc}
\tablecaption{LCE Fractions in [\sulf{2}]-weak Galaxies vs. Their Twins\label{tab:lyc_frac}}
\tablehead{
    \colhead{} &
    \colhead{LCE Fraction\tablenotemark{a}} &
    \colhead{Sig.\tablenotemark{b}} &
    \colhead{Strong LCE Fraction} &
    \colhead{Sig.}
   }
\startdata
$f_{\rm esc}({\rm Ly}{\alpha})$ & 0.59 $\pm$ 0.09 & 2.34 & 0.18 $\pm$ 0.09 & 2.97 \\
$[$\oxy{3}$]$/$[$\oxy{2}$]$ & 0.62 $\pm$ 0.10 & 2.00 & 0.15 $\pm$ 0.10 & 3.02 \\
$[$\oxy{1}$]$/H$\beta$ & 0.62 $\pm$ 0.12 & 1.74 & 0.31 $\pm$ 0.12 & 1.78 \\
${\rm EW}({\rm H}{\beta})$ & 0.50 $\pm$ 0.11 & 2.76 & 0.21 $\pm$ 0.11 & 2.48 \\
${\rm EW}({\rm Ly}{\alpha})$ & 0.71 $\pm$ 0.12 & 1.06 & 0.29 $\pm$ 0.12 & 1.91 \\
EW(\carb{2} 1334) & 0.47 $\pm$ 0.10 & 3.14 & 0.20 $\pm$ 0.10 & 2.66 \\
EW(\sili{2} 1260) & 0.35 $\pm$ 0.09 & 4.33 & 0.18 $\pm$ 0.09 & 2.97 \\
$M_{\rm FUV}$ & 0.54 $\pm$ 0.10 & 2.62 & 0.15 $\pm$ 0.10 & 3.02 \\
$M_{\rm NUV}$ & 0.40 $\pm$ 0.09 & 4.06 & 0.13 $\pm$ 0.09 & 3.35 \\
UV $\beta$ & 0.39 $\pm$ 0.10 & 3.88 & 0.22 $\pm$ 0.10 & 2.57 \\
$M_\star$ & 0.53 $\pm$ 0.09 & 2.89 & 0.13 $\pm$ 0.09 & 3.35 \\
$r_{50}$ & 0.56 $\pm$ 0.12 & 2.20 & 0.31 $\pm$ 0.12 & 1.78 \\
SFR$_{\rm H\beta}$/area & 0.47 $\pm$ 0.10 & 3.14 & 0.20 $\pm$ 0.10 & 2.66 \\
SFR$_{\rm UV}$/area & 0.59 $\pm$ 0.09 & 2.34 & 0.18 $\pm$ 0.09 & 2.97 \\
${\rm SFR_{\rm H\beta}}/M_\star$ & 0.50 $\pm$ 0.11 & 2.78 & 0.25 $\pm$ 0.11 & 2.26 \\
${\rm SFR_{\rm UV}}/M_\star$ & 0.47 $\pm$ 0.09 & 3.47 & 0.13 $\pm$ 0.09 & 3.35 \\
\enddata
\tablenotetext{a}{Pairs of [\sulf{2}]-weak (\ds\ $ \leq -0.2$) and non-[\sulf{2}]-weak galaxies sharing similar values of each parameter listed in column 1 are selected, and this column lists the LCE fractions in non-[\sulf{2}]-weak samples. These can be compared to the LCE fraction in the [\sulf{2}]-weak sample of $0.86 \pm 0.07$, and to the strong LCE of $0.59 \pm 0.10$.}
\tablenotetext{b}{Significance in the difference between a LCE fraction in the [\sulf{2}]-weak sample and that in the non-[\sulf{2}]-weak sample.}
\tablecaption{Uncertainties and statistical significances are estimated assuming binomial distributions.}
\end{deluxetable*}

\section{Discussion\label{sec:discuss}}

\subsection{Implications for photoionization models}

Fundamentally, there are two different mechanisms by which LyC leakage can occur from a star-forming region---a density-bounded nebula and a radiation-bounded nebula with holes \citep{Zackrisson2013}. The former refers to a scenario in which regions undergoing intense stellar formation fully ionize their surroundings and have ionizing photons leftover that can escape, while the latter refers to a ``picket-fence'' scenario in which supernovae or stellar winds clear out low-density channels in the neutral ISM, through which LyC photons escape \citep{Bergvall2006}.

 The density-bounded case is the simplest picture explaining the correlation between weak [\sulf{2}] and LyC escape, as the absence of \hy{1} near the edge of a Str\"{o}mgren sphere leads to a significant decrease in [\sulf{2}] emissions. However, a simple density-bounded model cannot account for the majority of LyC detections in UV absorption-line studies \citep{Chisholm2018,Gazagnes2018}, and for the observed optical emission-line ratios \citep{Ramambason2020}.

The ``picket-fence'' model therefore seems to be favored by observations. In this case, [\sulf{2}]-deficiency indicates that a significant fraction of the total solid angle as seen from the star-forming region is optically thin in the LyC. We also note that a reduced column density of the neutral gas in the fence is required to explain the largest observed \fesc\ \citep{Gazagnes2020,Ramambason2020}.

Instead of the above two classical one-zone models, \cite{Ramambason2020} propose two-zone photoionization models, in which ionization parameters and covering fractions are varied along different lines of sight. The two-zone models successfully reproduce the observed emission-line properties in low-$z$ LCEs, including [\sulf{2}]-deficiency, and are compatible to results from UV absorption-line studies.

\begin{figure*}
\gridline{
    \fig{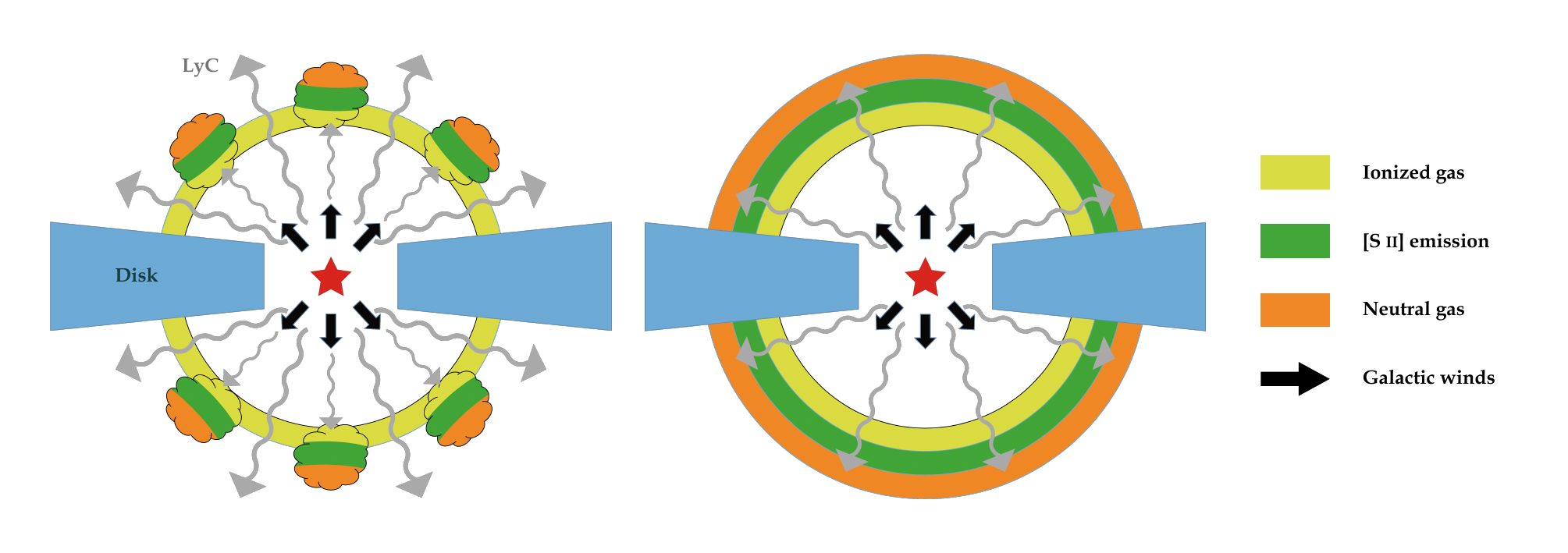}{1.0\textwidth}{}
}
\caption{Schematic diagrams for a LCE (left panel) and a non-LCE (right panel) in the context of a simple wind-blown shell. In the case of a LCE, only the optically thick clouds contribute to [\sulf{2}] emission while both the clouds and the optically thin regions between the clouds contribute to emission from Balmer and high ionization metal lines. In the case of a non-LCE, the entire shell contributes to the [\sulf{2}] emission. We note that they only represent the simplest cases where isotropy is assumed.\label{fig:cartoon}}
\end{figure*}

In Figure \ref{fig:cartoon} we show schematic diagrams of such a two-zone model for a LCE and a non-LCE in the context of a simple windblown shell. In the former case, only the optically thick clouds contribute to [\sulf{2}] emission while both the clouds and the optically thin regions between the clouds contribute to emission from Balmer and high-ionization metal lines. In the latter case the entire shell contributes to the [\sulf{2}] emission.

\subsection{Implications for escape fractions\label{sec:fesc}}
\begin{figure}
\gridline{
    \fig{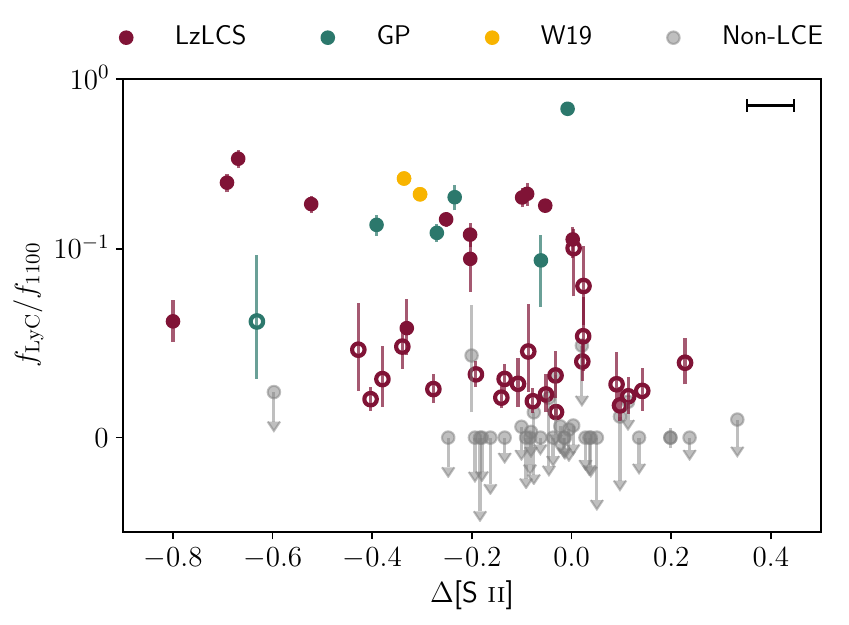}{0.47\textwidth}{(a)}
}
\gridline{
    \fig{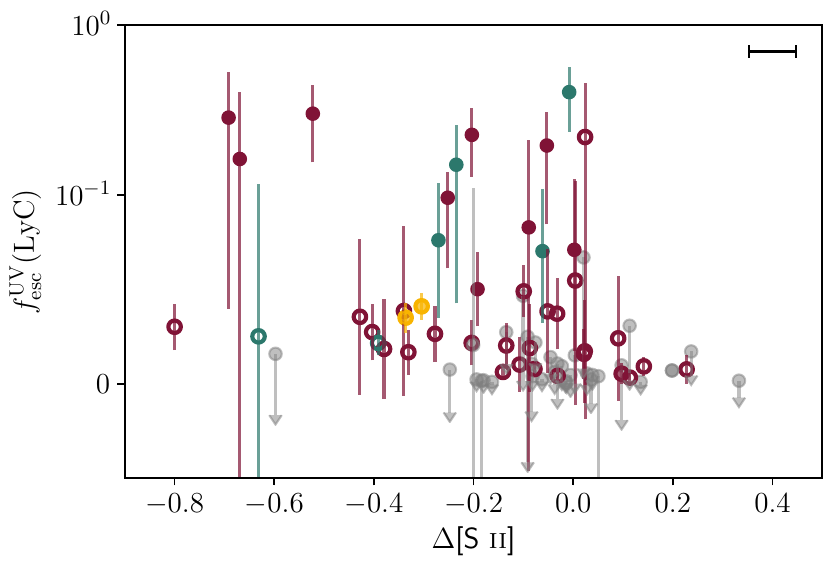}{0.47\textwidth}{(b)}
}
\caption{Escape fractions are plotted as functions of [\sulf{2}]-deficiency. A linear scale is used for \fesc\ $< 0.1$. Colored dots represent strong LCEs, circles represent weak LCEs, and gray dots represent upper limits. Most of the [\sulf{2}]-deficit galaxies (\ds\ $\lesssim -0.2$) are LCEs, but only weak (albeit statistically significant) correlations are found with \fesc\ (see Table \ref{tab:r}). \label{fig:fesc}}
\end{figure}

The relationship between \ds\ and \fesc\ shows substantial scatter, as evident in Figure \ref{fig:fesc}. Generalized Kendall's $\tau$ correlation coefficients which account for upper limits in \fesc\ are listed in Table \ref{tab:r} \citep{Isobe1986}. Our finding is in agreement with \cite{Ramambason2020}, who find that although [\sulf{2}]-deficiency can select LCE candidates, it is not obvious how to infer a numerical value for \fesc\ directly from \ds.

Several factors could contribute to the observed complexity. The simplest explanation is line-of-sight variations caused by porous \hy{2} regions. Observations from the Keck Lyman Continuum Spectroscopic Survey \citep{Steidel2018} and another sample of galaxies at $z \sim 3$ having high \oiiioii\ but low \fesc\ \citep{Nakajima2020} are both consistent with this picture. Using resolved stars to measure \fesc\ also shows that \fesc\ can vary significantly both due to the viewing angle and spatial resolution, even for the same galaxy \citep{Choi2020}.

A complication, though, arises from the possibility of anisotropically escaping LyC photons (e.g., \citealt{Zastrow2011,Cen2015ApJ}). This directly leads to ambiguity in the interpretation of non-detections---it is unclear if a galaxy not observed to have LyC escaping is truly a nondetection or if this is due to a particular orientation toward the observer. With a larger sample, we would be able to infer the fraction of LCEs with its proportional relationship to the \hy{1}\ covering fraction.

In short, an intricate interplay among factors dictates \fesc. Althoug unified models for describing different samples of LCEs have been proposed \citep{Cen2020}, it is likely that a combination of other properties (e.g., those shown in Figure 4 and with \mg{2} studied in \citealt{Henry2018,Chisholm2020}) is needed for accurately inferring \fesc.

In the face of this complexity, it seems that the determination of leakiness based on indirect signposts can only be done on a statistical basis rather than for individual objects (see \citealt{Runnholm2020} for predicting \lya\ radiation using multivariate regression). This reinforces the need for large samples, which was the main driver for the \lz. Even larger samples will ultimately be needed in the future.

\subsection{Analogs to high-$z$ galaxies}

Having discussed what can be learned from low-$z$ LCEs, this final subsection addresses whether they can truly represent the galaxies during the EoR. One question is the degree to which our definition of [\sulf{2}]-deficiency is based on galaxies at low $z$, which differ significantly from EoR galaxies. This can be addressed through spectroscopy using the James Webb Space Telescope (JWST) to construct BPT/VO diagrams like Figure \ref{fig:bpt} for EoR galaxies. It is promising that galaxies at $\langle z \rangle \sim$ 2.3 do follow the same locus as low-$z$ galaxies in Figure \ref{fig:bpt} \citep{Strom2018,Wang2019}.

In a recent analysis of simulated Lyman-break galaxies, \cite{Katz2020} claim that the sample of $z \sim 3$ LCEs \citep{Nakajima2020} are good analogs of EoR galaxies, while the [\sulf{2}]-deficient galaxies in W19 may not necessarily be so. The argument is largely based on whether the analogs populate the same regions in respective BPT/VO diagnostic diagrams. \cite{Katz2020} find that the LCEs in their simulations have deficits both in [\sulf{2}] and [\oxy{3}], indicating that the dominant effect is metallicity or mass rather than a property of the ISM.

We agree that metallicity or mass possibly play important roles \citep{Jaskot2019}, especially given that many LCEs fall at the top (metal-poor) end of the BPT/VO diagram as shown in Figure \ref{fig:bpt}. It is also true that galaxies are intrinsically complex systems, and thus it is difficult to disentangle primary and secondary correlations. The issue of whether high-$z$ and local galaxies can be analogs to one another is further complicated by the finding based on the MOSDEF–LRIS Survey that a similarity in the location of high-$z$ and local galaxies in the BPT/VO diagrams may not always be indicative of a similarity in their physical properties \citep{Topping2020}.

In general, while simulations shed light on the study of EoR galaxies, the multiphase nature of the interstellar and circumgalactic media as well as the high spatial resolution needed to explicitly simulate the LyC-escaping process reinforce the need for robust observational evidence. Given the relatively small pool of LCEs, the different physical properties exhibited by some of the [\sulf{2}]-deficient LCEs nevertheless offer valuable insights into the possibly different ways in which LyC can escape from galaxies.

\section{Conclusions\label{sec:conclu}}

We have reported on using the relative weakness of the [\sulf{2}] 6717, 6731 nebular emission lines defined with respect to normal star-forming galaxies as an indicator for galaxies that are optically thin to ionizing radiation. This method was proposed in \cite{Wang2019}, and statistically tested in this paper with new HST/COS observations of 66 star-forming galaxies in \lz\ \citep{lyc_sample}. We find that [\sulf{2}]-deficiency is an effective way to identify candidates for LyC-emitting galaxies, and can complement other proposed LyC predictors.

More specifically, we have shown that the LyC-emitting galaxies are more [\sulf{2}]-deficient than the other galaxies, that the detection fraction of them increases strongly as a function of [\sulf{2}]-deficiency, and that the value of the far-UV based escape fractions have statistically significant (but weak) correlations with [\sulf{2}]-deficiency. In addition, we have also shown that [\sulf{2}]-deficiency does not show a significant correlation with most of the other proposed indirect signposts of LyC leakage with the exception of the compactness of the starburst. This implies that [\sulf{2}]-deficiency is an independent indicator of LyC leakage.

We have discussed the photoionizing process in light of the scatter seen in the relationship between [\sulf{2}]-deficiency and \fesc. This likely indicates line-of-sight variations in ionization parameters and covering fractions, and/or anisotropically escaping LyC photons, which is in agreement with several other studies at low to intermediate redshifts \citep{Steidel2018,Ramambason2020}.

To summarize, the increased sample size of low-$z$ LyC-emitting galaxies from \lz\ allows us to statistically confirm that [\sulf{2}]-deficiency is a robust technique for finding galaxies leaking a significant amount of the LyC radiation. Although it is yet not obvious how accurately the numerical value of the escape fraction of LyC could be inferred from [\sulf{2}]-deficiency for individual galaxies, it is very useful in the context of statistical estimates for samples of galaxies. This gives us an additional technique to identify potential LyC-emitting galaxies at low $z$ and during the EoR with future observations with the JWST.

\acknowledgments
{
B.W. thanks Yiwei Sun for help making the schematic diagrams.
This work is supported by HST-GO-15626, provided by NASA through a grant from the Space Telescope Science Institute, which is operated by the Association of Universities for Research in Astronomy, Inc., under NASA contract NAS5-26555.
This publication made use of the NASA Astrophysical Data System for bibliographic information.

This project also made use of SDSS data. Funding for the Sloan Digital Sky Survey IV has been provided by the Alfred P. Sloan Foundation, the U.S. Department of Energy Office of Science, and the Participating Institutions. SDSS-IV acknowledges support and resources from the Center for High-Performance Computing at the University of Utah. The SDSS web site is www.sdss.org.
SDSS-IV is managed by the Astrophysical Research Consortium for the
Participating Institutions of the SDSS Collaboration including the
Brazilian Participation Group, the Carnegie Institution for Science,
Carnegie Mellon University, the Chilean Participation Group, the French Participation Group, Harvard-Smithsonian Center for Astrophysics,
Instituto de Astrof\'isica de Canarias, The Johns Hopkins University, Kavli Institute for the Physics and Mathematics of the Universe (IPMU) /
University of Tokyo, the Korean Participation Group, Lawrence Berkeley National Laboratory,
Leibniz Institut f\"ur Astrophysik Potsdam (AIP),
Max-Planck-Institut f\"ur Astronomie (MPIA Heidelberg),
Max-Planck-Institut f\"ur Astrophysik (MPA Garching),
Max-Planck-Institut f\"ur Extraterrestrische Physik (MPE),
National Astronomical Observatories of China, New Mexico State University,
New York University, University of Notre Dame,
Observat\'ario Nacional / MCTI, The Ohio State University,
Pennsylvania State University, Shanghai Astronomical Observatory,
United Kingdom Participation Group,
Universidad Nacional Aut\'onoma de M\'exico, University of Arizona,
University of Colorado Boulder, University of Oxford, University of Portsmouth,
University of Utah, University of Virginia, University of Washington, University of Wisconsin,
Vanderbilt University, and Yale University.
}

\,

\facilities{HST (COS), Sloan}
\software{Astropy v. 3.2.2 \citep{astropy2013,astropy2018}, Matplotlib v. 3.1.1 \citep{Matplotlib}, NumPy v. 1.17.2 \citep{numpy}, Pymccorrelation v 0.2.3 \citep{Curran2014,Privon2020}, SciPy v. 1.3.1 \citep{scipy}}

\newpage

\appendix
Here we list the LCE classifications of all galaxies in \lz, the remeasured SDSS [\sulf{2}] flux, and [\sulf{2}]-deficiency in Table \ref{tab:lyc_anci}. The scatter plots in Figure \ref{fig:other_cor} show correlations between \ds\ and other galaxy characteristics in the same way as Figure \ref{fig:ds2_cor}. The scattering is substantial in all cases.

\startlongtable
\begin{deluxetable}{lccc}
\tablecolumns{4}
\tablewidth{0pc}
\tablecaption{Flux Measurements of Galaxies in \lz. \label{tab:lyc_anci}}
\tablehead{
    \colhead{Galaxy} &
    \colhead{LCE Type\tablenotemark{a}} &
    \colhead{[\sulf{2}]} &
    \colhead{\ds}\\
    \colhead{} &
    \colhead{} &
    \colhead{($\times 10^{-17} {\rm erg \, cm^{-2} \, s^{-1} \, \mAA^{-1}}$)}&
    \colhead{(dex)}
    }
\startdata
J0036 & $\times$ & 8.17 $\pm$ 4.44 & -0.60 $\pm$ 0.24 \\
J0047 & $\circ$ & 51.06 $\pm$ 4.63 & 0.02 $\pm$ 0.04 \\
J0113 & $\circ$ & 45.98 $\pm$ 10.24 & 0.23 $\pm$ 0.10 \\
J0122 & $\circ$ & 11.81 $\pm$ 3.82 & -0.34 $\pm$ 0.14 \\
J0129 & $\times$ & 98.92 $\pm$ 10.38 & -0.10 $\pm$ 0.05 \\
J0723 & $\times$ & 54.75 $\pm$ 2.56 & 0.00 $\pm$ 0.02 \\
J0804 & $\bullet$ & 4.33\tablenotemark{$\dagger$} & -0.67   \\
J0811 & $\circ$ & 1.60\tablenotemark{$\dagger$} & -0.80   \\
J0814 & $\times$ & 207.83 $\pm$ 7.99 & -0.03 $\pm$ 0.02 \\
J0826 & $\times$ & 19.14 $\pm$ 5.77 & -0.25 $\pm$ 0.13 \\
J0834 & $\times$ & 104.46 $\pm$ 12.70 & -0.01 $\pm$ 0.05 \\
J0909 & $\bullet$ & 4.30   & -0.69   \\
J0911 & $\circ$ & 149.60 $\pm$ 13.13 & -0.20 $\pm$ 0.04 \\
J091207 & $\times$ & 124.31 $\pm$ 4.79 & -0.05 $\pm$ 0.02 \\
J091208 & $\times$ & 44.18 $\pm$ 4.50 & -0.08 $\pm$ 0.04 \\
J0917 & $\bullet$ & 79.66 $\pm$ 8.61 & -0.25 $\pm$ 0.05 \\
J0925 & $\times$ & 20.05 $\pm$ 29.71 & -0.08 $\pm$ 0.64 \\
J0933 & $\bullet$ & 11.55 $\pm$ 4.00 & -0.52 $\pm$ 0.15 \\
J0940 & $\times$ & 49.59 $\pm$ 5.64 & 0.03 $\pm$ 0.05 \\
J0952 & $\circ$ & 9.64 $\pm$ 7.02 & -0.40 $\pm$ 0.32 \\
J0957 & $\times$ & 302.97 $\pm$ 14.10 & -0.19 $\pm$ 0.02 \\
J0958 & $\circ$ & 13.43 $\pm$ 10.01 & -0.38 $\pm$ 0.32 \\
J1014 & $\times$ & 23.52 $\pm$ 3.09 & -0.18 $\pm$ 0.06 \\
J1026 & $\circ$ & 23.63   & 0.14   \\
J1033 & $\bullet$ & 72.51 $\pm$ 11.64 & -0.05 $\pm$ 0.07 \\
J1038 & $\circ$ & 158.05 $\pm$ 8.43 & -0.14 $\pm$ 0.02 \\
J1051 & $\times$ & 45.32 $\pm$ 4.46 & 0.04 $\pm$ 0.04 \\
J1053 & $\circ$ & 225.05 $\pm$ 15.12 & -0.08 $\pm$ 0.03 \\
J1055 & $\times$ & 36.27 $\pm$ 4.04 & 0.14 $\pm$ 0.05 \\
J1104 & $\times$ & 32.19 $\pm$ 4.98 & 0.11 $\pm$ 0.07 \\
J1122 & $\circ$ & 11.44 $\pm$ 5.24 & 0.00 $\pm$ 0.20 \\
J1128 & $\circ$ & 26.65 $\pm$ 3.34 & 0.11 $\pm$ 0.06 \\
J1129 & $\times$ & 15.49 $\pm$ 3.67 & -0.04 $\pm$ 0.11 \\
J1133 & $\circ$ & 41.98 $\pm$ 8.48 & -0.09 $\pm$ 0.09 \\
J1158 & $\circ$ & 218.07 $\pm$ 12.62 & -0.10 $\pm$ 0.03 \\
J1159 & $\circ$ & 8.38 $\pm$ 2.68 & -0.43 $\pm$ 0.14 \\
J1209 & $\times$ & 164.70 $\pm$ 6.13 & -0.08 $\pm$ 0.02 \\
J1219 & $\circ$ & 22.67   & -0.11   \\
J1235 & $\bullet$ & 29.19 $\pm$ 11.95 & -0.19 $\pm$ 0.18 \\
J1240 & $\times$ & 21.31   & -0.13   \\
J1244 & $\times$ & 264.26 $\pm$ 11.60 & 0.10 $\pm$ 0.02 \\
J1246 & $\circ$ & 87.90 $\pm$ 4.44 & 0.10 $\pm$ 0.02 \\
J1248 & $\circ$ & 108.88 $\pm$ 9.37 & 0.09 $\pm$ 0.04 \\
J1249 & $\times$ & 26.51 $\pm$ 4.13 & 0.05 $\pm$ 0.07 \\
J1255 & $\times$ & 53.52   & 0.24   \\
J1257 & $\times$ & 46.72 $\pm$ 4.23 & -0.02 $\pm$ 0.04 \\
J1300 & $\times$ & 6.14   & -0.18   \\
J1301 & $\circ$ & 55.34 $\pm$ 13.69 & -0.13 $\pm$ 0.11 \\
J1305 & $\bullet$ & 10.05   & -0.20   \\
J1310 & $\circ$ & 37.72 $\pm$ 8.34 & -0.33 $\pm$ 0.10 \\
J1314 & $\times$ & 178.42 $\pm$ 11.23 & -0.02 $\pm$ 0.03 \\
J1319 & $\times$ & 78.26 $\pm$ 7.72 & -0.16 $\pm$ 0.04 \\
J1326 & $\circ$ & 72.90 $\pm$ 5.34 & 0.02 $\pm$ 0.03 \\
J1329 & $\times$ & 145.64 $\pm$ 15.73 & -0.01 $\pm$ 0.05 \\
J1345 & $\times$ & 208.06 $\pm$ 11.61 & 0.04 $\pm$ 0.02 \\
J1350 & $\times$ & 39.36   & 0.33   \\
J1403 & $\circ$ & 27.32 $\pm$ 23.46 & -0.03 $\pm$ 0.37 \\
J1410 & $\bullet$ & 24.62 $\pm$ 2.44 & 0.00 $\pm$ 0.04 \\
J1440 & $\circ$ & 199.28 $\pm$ 11.96 & -0.03 $\pm$ 0.03 \\
J1517 & $\bullet$ & 78.45 $\pm$ 4.58 & -0.09 $\pm$ 0.03 \\
J1540 & $\times$ & 189.29 $\pm$ 12.16 & -0.06 $\pm$ 0.03 \\
J1559 & $\times$ & 90.14 $\pm$ 7.19 & 0.02 $\pm$ 0.03 \\
J1604 & $\times$ & 34.26 $\pm$ 7.46 & -0.09 $\pm$ 0.09 \\
J1646 & $\circ$ & 26.62 $\pm$ 11.65 & -0.28 $\pm$ 0.19 \\
J1648 & $\circ$ & 11.68 $\pm$ 2.90 & 0.02 $\pm$ 0.11 \\
J1720 & $\circ$ & 28.40   & -0.05   \\
\enddata
\tablenotetext{a}{Classes of LCEs: $\bullet$ strong LCEs, $\circ$ weak LCEs, $\times$ non-LCEs.}
\tablenotetext{\dagger}{Upper limit inferred from $3\sigma$ background.}
\end{deluxetable}

$\quad$

\vspace{44pt}

\begin{figure*}
\gridline{
    \fig{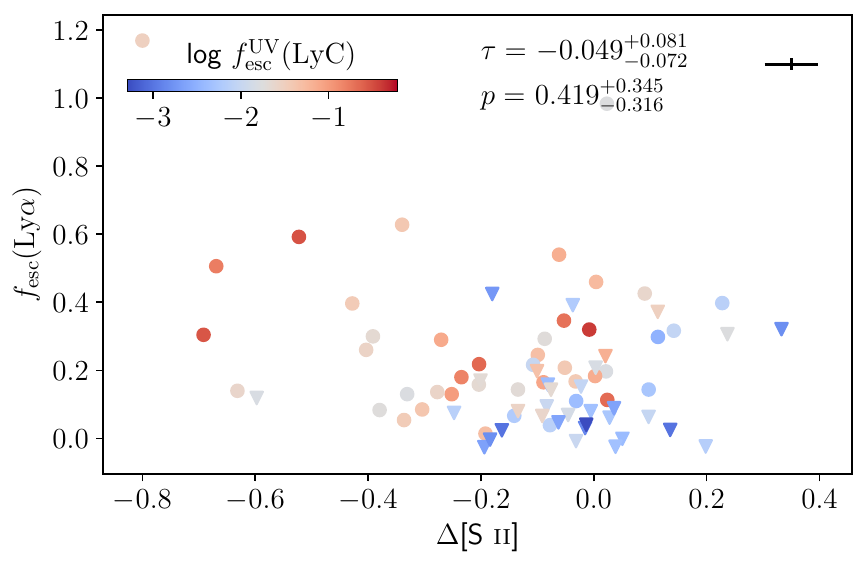}{0.40\textwidth}{}
    \fig{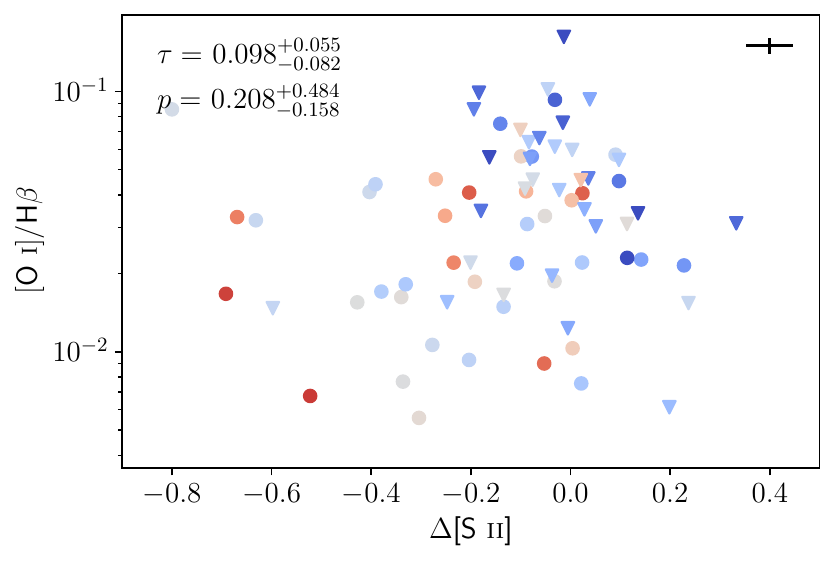}{0.40\textwidth}{}
}
\gridline{
    \fig{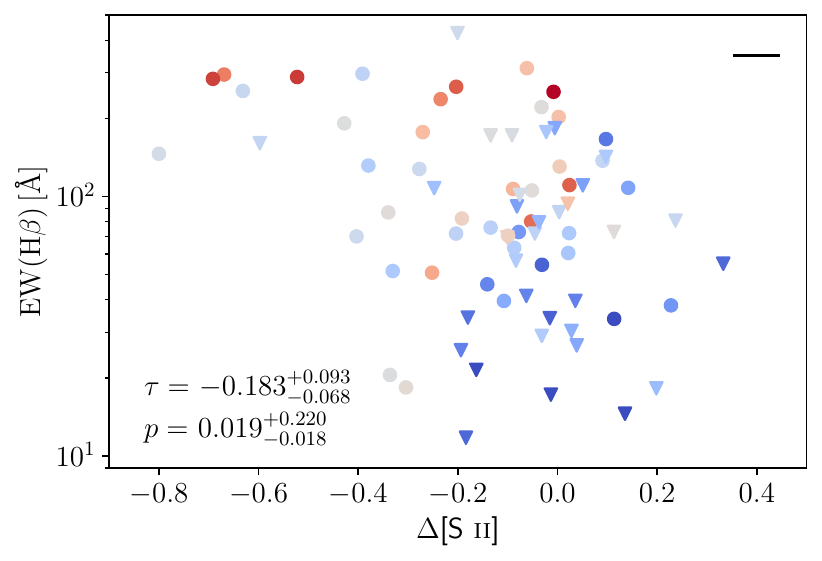}{0.40\textwidth}{}
    \fig{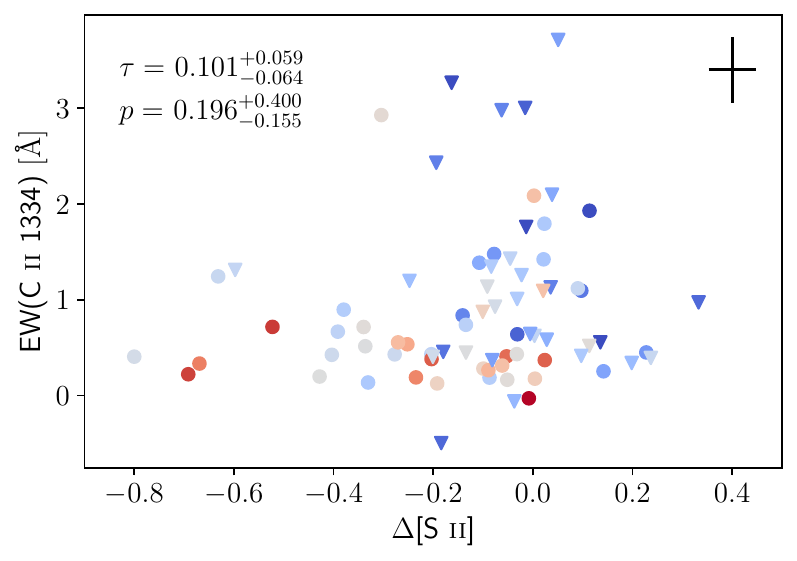}{0.40\textwidth}{}
}
\gridline{
    \fig{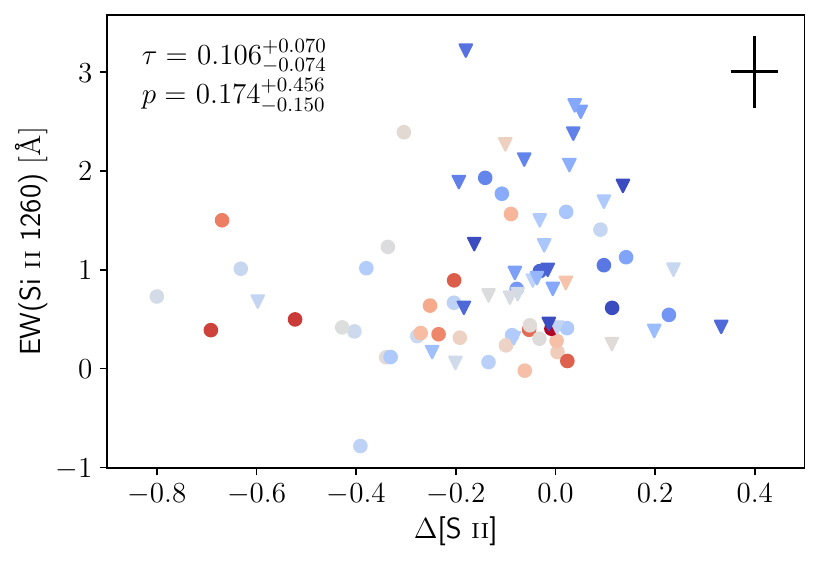}{0.40\textwidth}{}
    \fig{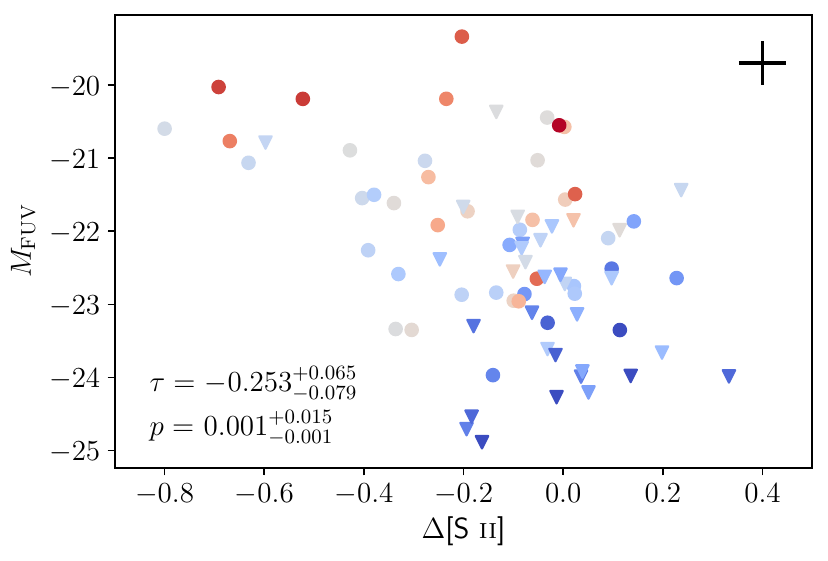}{0.40\textwidth}{}
}
\gridline{
    \fig{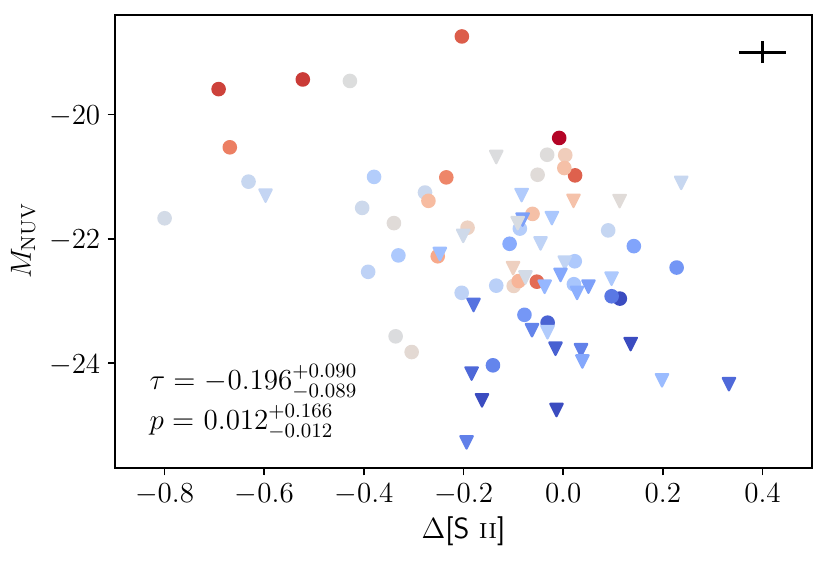}{0.40\textwidth}{}
    \fig{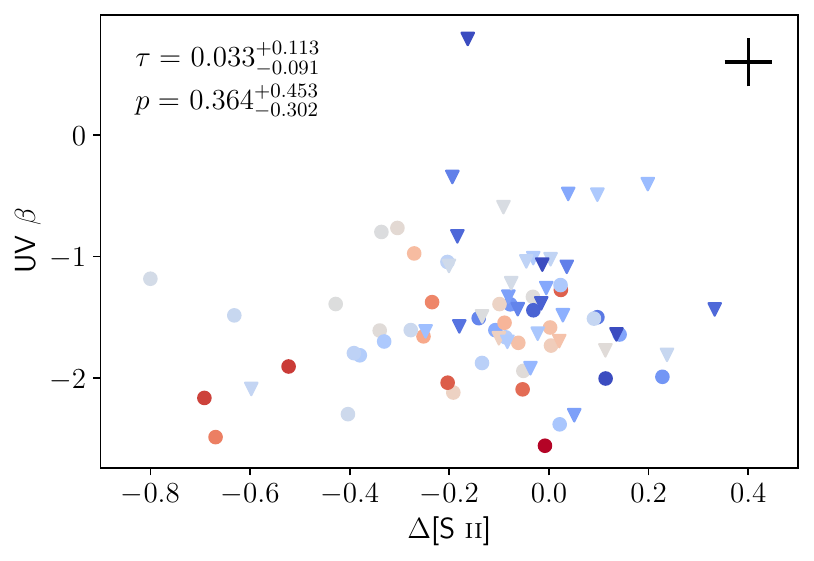}{0.40\textwidth}{}
}
\caption{Same as Figure \ref{fig:ds2_cor}, but with different galaxy characteristics. Colors indicate \fesc. Triangles indicate that the associated \fesc\ are upper limits. $\tau$ and $p$ are Kendall's correlation coefficient and the $p$-value, respectively.\label{fig:other_cor}}
\end{figure*}

\setcounter{figure}{5}
\begin{figure*}
\gridline{
    \fig{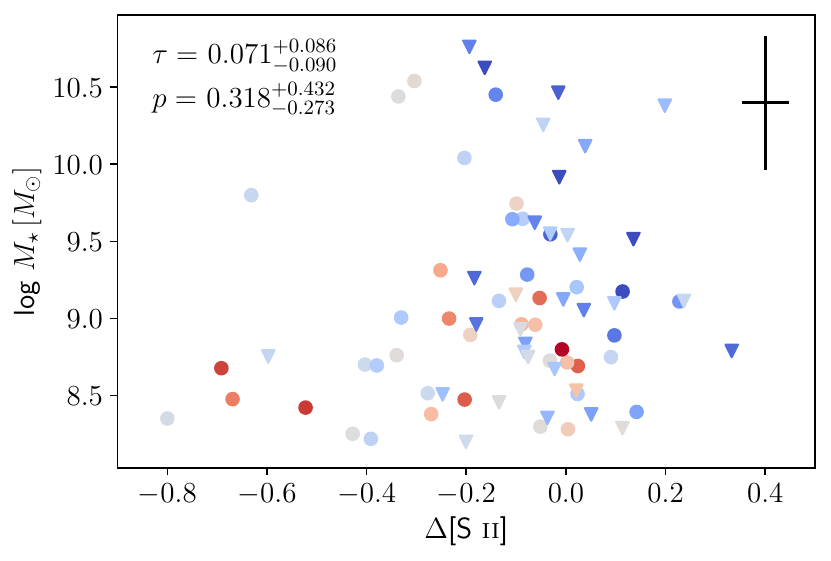}{0.40\textwidth}{}
    \fig{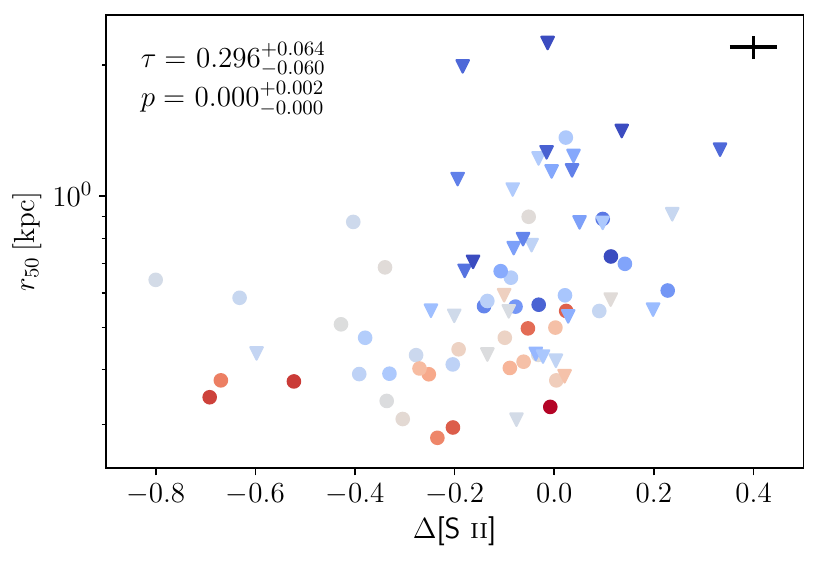}{0.40\textwidth}{}
}
\gridline{
    \fig{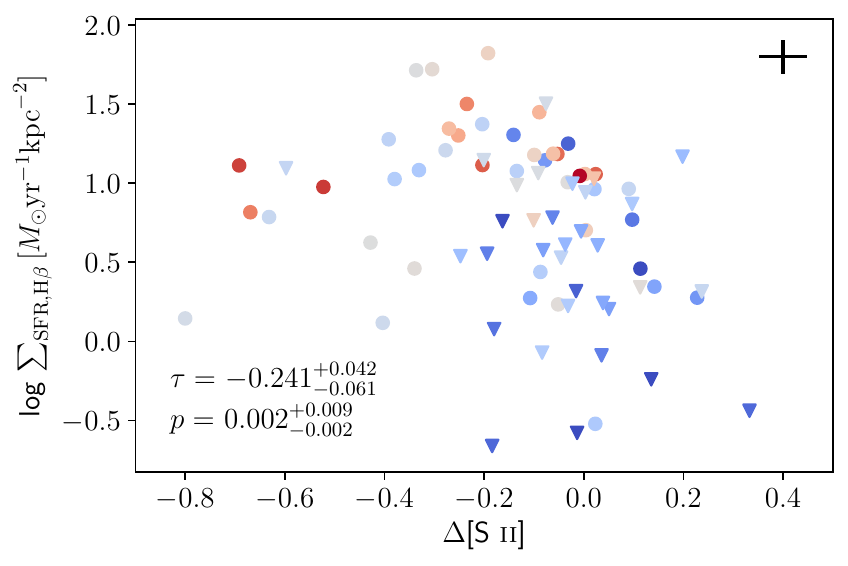}{0.40\textwidth}{}
    \fig{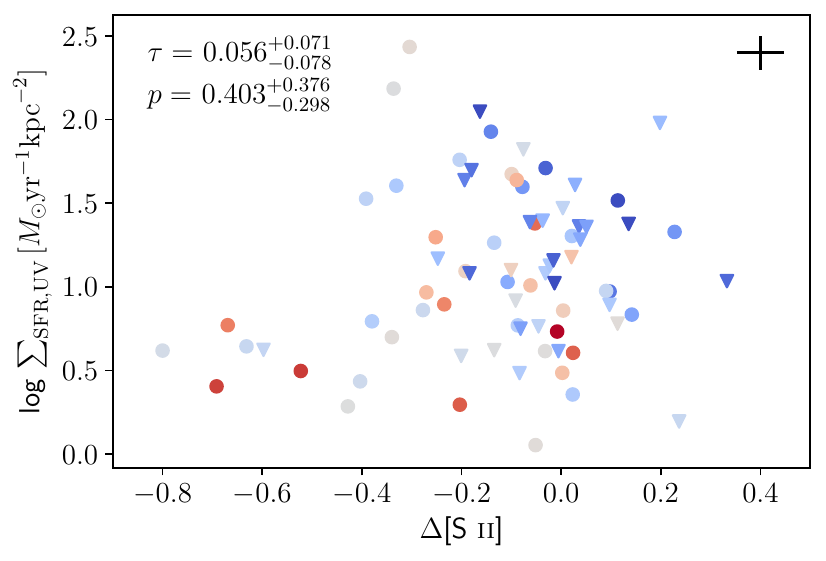}{0.40\textwidth}{}
}
\gridline{
    \fig{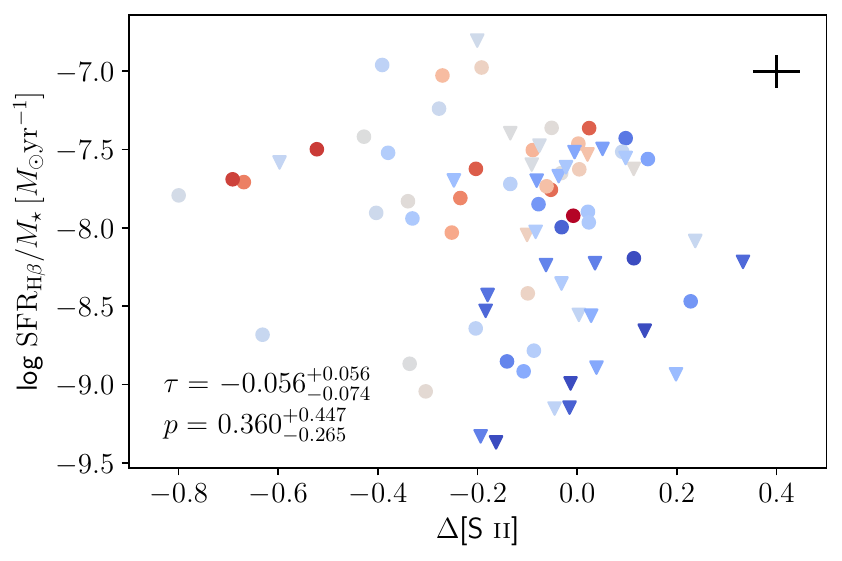}{0.40\textwidth}{}
    \fig{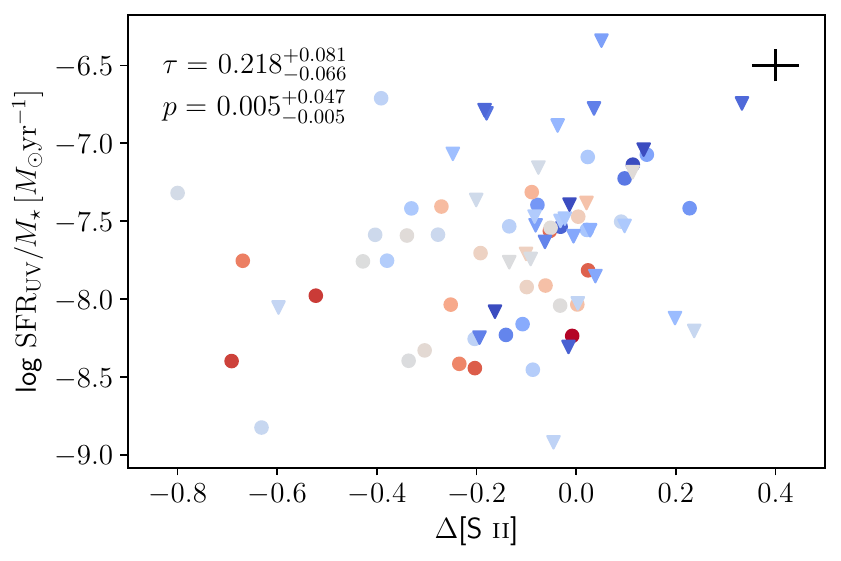}{0.40\textwidth}{}
}
\caption{(Continued.)}
\end{figure*}

\bibliography{lyc_sii.bib}

\end{document}